\documentclass[12pt]{iopart}

\usepackage[dvips]{graphicx}
\usepackage{colordvi}
\usepackage{subfigure}

\usepackage{color}
\definecolor{darkblue}{rgb}{0,0,0.6}
\definecolor{darkred}{rgb}{0.7,0,0}
\definecolor{darkgreen}{rgb}{0,0.7,0}
\usepackage{amsfonts}

\newcommand{\bea}{\begin{eqnarray}}
\newcommand{\eea}{\end{eqnarray}}
\newcommand{\be}{\begin{equation}}
\newcommand{\ee}{\end{equation}}

\begin{document}

\title{Nonequilibrium Response from the dissipative Liouville Equation}

\author{Matteo Colangeli\footnote{Present address:
Dipartimento di Matematica, Politecnico di Torino, Corso Duca degli Abruzzi 24, I-10129 Torino, Italy}}

\address{School of Mathematical Sciences, Queen Mary University of London, 
Mile End Road, London E1 4NS, UK}
\ead{colangeli@calvino.polito.it}

\begin{abstract}
The problem of response of nonequilibrium systems is currently under intense investigation. We propose a general method of solution of the Liouville Equation for thermostatted particle systems subjected to external forces which retains only the slow degrees of freedom, by projecting out the majority of fast variables. Response formulae, extending the Green-Kubo relations to dissipative dynamics are provided, and comparison with numerical data is presented.
\end{abstract}


\maketitle

\section{Introduction}
The celebrated theory of linear response of systems close to
equilibrium has been developed in the first half of the XX century,
and has proven highly successful in describing a very wide range of macroscopic
phenomena, particularly thanks to the usefulness of the Fluctuation-Dissipation Theorems (FDT)
\cite{CTXX}. In the second half of the XX century, numerous generalizations have been proposed
to extend the classical theory beyond the linear regime, by
considering the microscopic evolution as represented by deterministic
dynamical systems, by dynamical systems perturbed by some appropriate noise, or by purely
stochastic processes, which reach a nonequilibrium steady state. In particular, relations known as Fluctuation Relations
have been introduced together with different nonequilibrium generalizations of the FDT.
One then investigates the variations of these dynamics under certain perturbations. Remarkably, it turns out that rather different
approaches lead to similar results, cf.\ Ref.\cite{BPRVetc} for recent
reviews of the historical developments and of the recent results. The recent developments,
however, do not constitute a consistent and comprehensive theory yet, therefore a variety
of different approaches is worthy investigating.  For instance, Ruelle has shown \cite{CS07} that, for uniformly
hyperbolic dynamical systems, the linear response away from equilibrium is very similar to the linear response close to equilibrium: the Kramers-Kronig dispersion relations hold, and the FDT survives in a modified form, which accounts for the oscillations around
the relevant attractors. If the chaotic hypothesis does not hold, Ruelle concludes that
two new phenomena may arise. The first is a violation of linear response, in the sense that the nonequilibrium steady state does not depend differentiably on parameters. The second phenomenon is a violation of the dispersion relations: the susceptibility has singularities
in the upper half complex plane. These `acausal' singularities are actually due to `energy
nonconservation': for a small periodic perturbation of the system, the amplitude of the
linear response is arbitrarily large.
This means that the steady state of the dynamical system
under study is not `inert' but can give energy to the outside world, something rather
different from the behavior of an equilibrium state.
This approach is based on the smoothness property of SRB measures \cite{Young} along the unstable fibres of the attractors, something hardly realizable in the phase space probability distribution of physically relevant
nonequilibrium systems. Therefore, this approach is limited to the properties
of systems whose departure from the ideal uniform hyperbolicity does not seriously affect
the observed behavior \cite{CS07}.
From a different perspective, Vulpiani \textit{et al.} \cite{BLMV} have studied the response of dynamical systems
to finite amplitude perturbation, assuming that their phase space probability distributions
are not singular with respect to the Lebesgue measure, hence have an integrable density. These authors found that a generalized fluctuation-response relation holds, and it links the average
relaxation towards the steady state to the invariant measure of the system. Furthermore, this approach
points out the relevance of the amplitude of the initial perturbation, something which does not
pertain to the approach mentioned above. The use of regular distributions is justified by
considering that any physical phenomenon is affected by some noise, capable of smoothing out the
singularities produced by the dissipative deterministic dynamics.
However, one may also like to consider the noise as emerging from a projection procedure \cite{Zwanzig}, or may investigate the possibility that the distribution (not necessarily invariant) be expressed in terms of appropriate expansions around a reference one. In this work we pursue the latter approach and propose a method which resembles the Chapman-Enskog technique \cite{Chap} for the Boltzmann Equation (BE), which is suitable for the generalized Liouville Equation (LE) of deterministically thermostatted particle systems. The method takes inspiration from the Bogolyubov hypothesis \cite{Bog} of time scales separation and postulates that the dynamics of the single-particle distribution function is driven by the dynamics of a set of selected fields \cite{Matteo}. Analogously, in the context of many-particle systems, one can conjecture that the dynamics of the phase space probability density is triggered by the dynamics of some averaged phase space quantities.
In particular, in the theory of the BE, the notion of slow variables was corroborated by the assumption of local equilibrium.
At the level of the LE, on the other hand, a similar notion can be introduced under the assumption of a weakly dissipative dynamics, in order to let the resulting probability density be not too far from the Gibbsian density, which is required for the application of the method.

\section{The Chapman-Enskog method of solving the Boltzmann Equation}
\label{sec:sec2}

 In this Section we review some of the main features of the Chapman-Enskog method, which we intend, next, to employ, in Sec.\ref{sec:sec3}, in the context of thermostatted particle systems, to study the evolution of the probability density in the full phase space. The approach, introduced by Enskog and popularized in a modified version by Chapman and Cowling \cite{Chap}, consists of a mathematical procedure meant to approximate the solution of the BE. The method deserves a special mention in kinetic theory of gases since the Boltzmann pioneering works, as it allows to derive the Navier-Stokes-Fourier equations of hydrodynamics from the BE, by elucidating and making use of concepts, such as the scaling (hydrodynamic) limit of a kinetic equation, the notion of local equilibrium and the time scales separation, which led to the systematic construction of projection operator methods in statistical mechanics  as well as to the later development of special large deviation methods in the theory of stochastic processes \cite{Jona}.
Let us start by writing down the BE, which (in absence of external forces) reads:

\be
\partial_{t}f=-\textbf{v} \cdot \nabla f + C(f) \label{BE}
\ee

where $f=f(\textbf{r},\textbf{v},t)$ is the one particle distribution function, depending on position $\textbf{r}$, particle velocity $\textbf{v}$ and time $t$ and $C(f)$ is a nonlinear integral collision operator. By rescaling the BE with characteristic time and length scales, the Knudsen number $\epsilon=\frac{\lambda}{L}$ arises naturally from (\ref{BE}), which, in dimensionless form, becomes:

\be
\partial_{t}f=-\textbf{v} \cdot \nabla f + \frac{1}{\epsilon}C(f) \label{BECE}
\ee

 where $L$ denotes a macroscopic length scale (e.g. the sice of the system) and $\lambda$ is the \textit{mean free path}, i.e. the average distance covered by a moving particle between successive collisions with other moving particles.
In the limit $\epsilon\rightarrow0$, which is commonly referred to as the \textit{hydrodynamic limit}, the fluid becomes dense enough that the dynamics, described by (\ref{BECE}), is dominated by the effect of the collisions.
 By introducing the notion of \textit{local equilibrium}, (i.e. the existence of macroscopic regions each of which is in an equilibrium state which may be different from the equilibrium states found in other regions) it is possible to define, locally, a set of selected variables $\textbf{x}$, known as \textit{hydrodynamic fields} (which correspond to the collision invariants: number of particles density $n(\textbf{r},t)$, momentum density $n(\textbf{r},t)\textbf{u}(\textbf{r},t)$, and kinetic energy density $e(\textbf{r},t)=\frac{3}{2}n(\textbf{r},t) k_B T(\textbf{r},t)$), which enter the definition of Maxwell-Boltzmann statistics:

 \be
f^{LM}(\textbf{r},\textbf{v},t)=n(\textbf{r},t) (\frac{m}{2 \pi
k_{B}T(\textbf{r},t)})^{\frac{3}{2}}e^{-\frac{m(\textbf{v}-\textbf{u}(\textbf{r},t))^{2}}{2
k_{B}T(\textbf{r},t)}} \label{LM} \ee

where $m$ is the mass of the particle. The local Maxwellians (\ref{LM}) are not, in general, solutions of (\ref{BE}). Nevertheless, in the hydrodynamic limit, where the length scale of the spacial inhomogeneities, $L\sim \epsilon^{-1}$, tends to diverge, they offer a good approximation to the exact solution. In fact, it is evident from Eq. (\ref{BECE}) that for $\epsilon\rightarrow0$ the last term on the right becomes singular. Then, the only way to avoid the singularity is that the collision term itself vanishes, which is guaranteed by the form of the local Maxwellians.
It is worth to notice that Eq. (\ref{BECE}) can also be considered, in the Fourier-Laplace space, as an eigenvalue problem for the operator $\Lambda=C(f)-i \epsilon\textbf{k} \cdot\textbf{v}$.  As discussed in Ref. \cite{Blatt}, some technical problems arise when one attempts to solve the eigenvalue problem away from the strict hydrodynamic limit $\epsilon\rightarrow0$, and alternative techniques, based on projection operators, have been recently made available \cite{Matteo2}.
However, in those circumstances where $\epsilon \ll 1$, it makes sense to attempt a perturbative method to solve Eq. (\ref{BECE}), by introducing the following expansion:

\be
f=\sum_{l=0}^{\infty}\epsilon^{l} f^{(l)}\label{fexp}
\ee
Thus, by inserting (\ref{fexp}) into (\ref{BECE}), one may define a \textit{microscopic} time derivative $\partial_t^{micro}f$ as:

\be
\partial_t^{micro}f =-\textbf{v} \cdot \nabla[\sum_{l=0}^{\infty} f^{(l)}\epsilon^{l}]+\frac{1}{\epsilon}C(\sum_{l=0}^{\infty} f^{(l)}\epsilon^{l}) \label{microBE}
\ee

Following \cite{Matteo2}, in the theory of the BE it is also possible to introduce a \textit{macroscopic} time derivative, $\partial_t^{macro}f$, provided that the assumption of local equilibrium holds. The idea is to employ the notion of \textit{normal solutions}, which allows to express the spatial and temporal dependence of the terms $f^{(l)}$ through the hydrodynamic fields, $f^{(l)}(\textbf{r},\textbf{v},t)=f^{(l)}(\textbf{x}(\textbf{r},t),\textbf{v})$.
By applying a sort of chain rule, we can, then, write:

\be
\partial_t^{macro}f =\frac{\partial (\sum_{j=0}^{\infty}\epsilon^j f^{(j)})}{\partial \textbf{x}}\partial_{t}\textbf{x} \label{macroBE}
\ee
The time derivative of the hydrodynamics fields, in (\ref{macroBE}), is found to be given by:
\be
\partial_{t}\textbf{x}=\sum_{k=0}^{\infty}\epsilon^{k}\partial_{t}^{(k)}\textbf{x} \label{dtx}
\ee

where $\partial_{t}^{(k)}\textbf{x}=-\int \textbf{w}(\textbf{v})\textbf{v}\cdot \nabla f^{(k)} d^{3}\textbf{v}$ and where  the $\textbf{w}$'s are lower order Sonine polynomials of $\textbf{v}$.
%obtain the expression for the macroscopic time derivative, given by:
%\be
%&=&\frac{\partial [\sum_{k=0}^{\infty}\epsilon^k f^{(k)}]}{\partial x_{\alpha}}\partial_{t}[\int w_{\alpha}(\textbf{v}) %\sum_{j=0}^{\infty}\epsilon^j f^{(j)}d^{3}\textbf{v}] \nonumber\\
%\partial_t^{macro}f =\sum_{j,k=0}^{\infty}\epsilon^{j+k}\frac{\partial f^{(k)}}{\partial \textbf{x}} \partial_{t}^{(j)}\textbf{x}= \sum_{j,k=0}^{\infty}\epsilon^{j+k}\partial_{t}^{(j)}f^{(k)} \label{macroBE}
%\ee
The method, then, requires to equalize the two time derivatives $\partial_t^{micro}f$ and $\partial_t^{macro}f$ (supplemented by (\ref{dtx})). The equality, in fact, provides a cascade of equations for the various terms of the expansion $f^{(l)}(\textbf{x}(\textbf{r},t),\textbf{v})$, whose leading order is represented by the family of Maxwellians (\ref{LM}), and whose first correction, $f^{(1)}$, can be obtained \cite{Chap} from:
%Going to the next order, $\epsilon^{0}$, the following equation is obtained for $f^{(1)}$:

%\be
%\partial_{t}^{(0)}f^{(0)}=-\textbf{v} \cdot \nabla f^{(0)}+ C(f^{(1)}) \nonumber
%\ee

%which, with (\ref{macroBE}) and (\ref{derivativeCE}), becomes:

\be
\frac{\partial f^{(0)}}{\partial \textbf{x}}\int \textbf{w}(\textbf{v}) \textbf{v} \cdot \nabla f^{(0)}(v) d^{3}\textbf{v}= \textbf{v} \cdot \nabla f^{(0)}- C(f^{(1)}) \label{1storder}
\ee

The lowest order of the equations of hydrodynamics (\ref{dtx}) is known as \textit{Euler} hydrodynamics and features vanishing transport coefficients, whereas, by adding the first correction $f^{(1)}$, one obtains the dissipative Navier-Stokes-Fourier equations, which are endowed with constitutive expressions for the stress tensor and the heat flux \cite{Matteo2}.
%:
%\bea
%\mbf{\sigma}&=&\frac{1}{3}Tr[\mbf{\sigma}]\textbf{I}+\mbf{\sigma}^{(s)}=-\zeta
%(\nabla\cdot\textbf{u})\textbf{I}-2\eta[\nabla \textbf{u}]^{s} \label{NS}\\
%\q&=&-\lambda \nabla T \label{F}
%\eea
%where $\mbf{\sigma}^{(s)}$ denotes
%the symmetric traceless part of the stress tensor and $\zeta, \eta,
%\lambda$ are transport coefficients called, respectively, bulk
%viscosity, shear viscosity and thermal conductivity.
However, higher order corrections of the Chapman-Enskog method,
resulting in hydrodynamic equations with higher derivatives (Burnett
and super-Burnett hydrodynamic equations) face severe difficulties
both from theoretical, as well as from the practical point of view
and various regularization methods have been suggested \cite{Matteo,Bobyl}.

\section{Dissipative Liouville Equation with external forcing and nonequilibrium response}
\label{sec:sec3}

The method discussed in the previous Section and applied in the context of kinetic theory can
be extended to Hamiltonian particle systems, as it is formally possible to
construct perturbative techniques to solve the LE. Such perturbation theories allow
to compute the higher order corrections to the phase space counterpart of the \textit{local equilibrium} in the BE and lead to expressions of transport coefficients in terms of the detailed microscopic dynamics.
The standard method which succeeds in this derivation was originally proposed by Green and Kubo \cite{GK}. Any perturbation theory based on an expansion of the probability density in powers of a small parameter $\epsilon$ needs to extend and justify the assumption of \textit{local equilibrium} in phase space and to prove that higher order corrections are suitable to refine the lower order approximations. This is a delicate issue, as, formally, there is no indication of the convergence of such an expansion in phase space and, mostly, there is no analog of the Caflish Theorem \cite{Caf}, formally proving that a proper truncation of the series (\ref{fexp}) approximates the solution to the BE. Furthermore, as pointed out by the Authors in \cite{Marra}, in order to obtain finite transport coefficients, one needs to postulate a strong decay property of the time autocorrelations of mass, momentum and energy currents.
We intend to show, here, that, in the context of externally driven thermostatted particles systems, the analogy with kinetic theory can be, to some extent, retained and most of the concepts formerly employed in the derivation of hydrodynamics from the BE, can be used to obtain nonequilibrium response formulae from the LE. In particular, the dimensionless parameter $\epsilon$ which arises after a proper rescaling of the LE, does not constitute just a mathematical book-keeping device which is eventually set to unity, but it can be endowed with a physical content, such as it was in the case of the BE.
We consider a particle system, held, for time $t\in (-\infty,0]$, in equilibrium with an external bath at temperature $T$.
At time $t=0$ an external force $\textbf{F}^{ext}$, starts acting and, hence, some energy is ''pumped'' inside the system. At the same time a deterministic thermostat is switched on in order to remove part (or the whole) of the energy provided by the external force, and to achieve, on a large time scale, a nonequilibrium steady state. The dynamics we have in mind are the deterministic thermostatted dynamics of nonequilibrium molecular dynamics \cite{EM}.
In particular, we refer to those systems for which the dimensionless LE takes the form:

\be
\dot{\rho}= -\dot{\Gamma} \cdot\nabla \rho + \epsilon \kappa \rho
\label{Liouv}
\ee
with:
\be
\dot{\Gamma}=\dot{\Gamma}_{0}+\epsilon R(\Gamma)=S \cdot \nabla H_{0}+\epsilon R(\Gamma) \label{eom}
\ee
where $S$ is the symplectic matrix, $H_{0}$ is the Hamiltonian of the many-particle system, $\dot{\Gamma}_{0}$ denotes the corresponding Hamiltonian contribution to the dynamics and $R(\Gamma)$ is the term which spoils the conservativity of the equations of motion. Furthermore, in (\ref{Liouv}), we introduced a shorthand notation for the phase space contraction rate: $\kappa=- \nabla \cdot \dot{\Gamma}$ and $\epsilon$ is a dimensionless parameter which is proportional to the intensity of the external driving.
A paradigmatic example, where the parameter $\epsilon$ attains a sensible structure, is provided by the Gaussian thermostatted Lorentz gas of hard spheres, with Hamiltonian $H_{0}(\Gamma)=\frac{\textbf{p}^{2}}{2 m}$. By adding an external force, here represented by $\mathbf{F}^{ext}=q \mathbf{E}$, with $q$ the electric charge and $\textbf{E}$ the external electric field, and the thermostat, both entering the definition of $R$, in (\ref{eom}), one finds: $R =q \textbf{E}-\alpha(\textbf{p})\textbf{p}$, where $m$ the mass of the particle and $\alpha=\frac{q \textbf{E}\cdot \textbf{p}}{p^{2}}=\kappa(\Gamma)$. For simplicity, we consider just a single particle and we also do not take the presence of scatterers explicitly into account, as it does not contribute to the phase space contraction rate, which, as discussed below, is the only observable we are going to be concerned with, in our model.
By rescaling all dimensional quantities with proper characteristic scales, the resulting dimensionless LE takes precisely the structure indicated in (\ref{Liouv}), and, in analogy with the definition of the Knudsen number introduced in Sec. \ref{sec:sec2}, the parameter $\epsilon$ is, then, given by the ratio of the energy absorbed by the thermostat, $\Delta\mathcal{E}_{diss}$, to a characteristic energy of the system $K$:

\be
\epsilon=\frac{E L}{v^{2}}\sim\frac{\Delta\mathcal{E}_{diss}}{K} \label{eps}
\ee

where denotes $E$ a typical intensity of the applied electric field, $L$ a reference length scale, $v$ a characteristic velocity of the dynamics and $K$ may be identified by the kinetic energy of the system or, in systems enjoying local thermodynamic equilibrium, may be also related to the temperature $T$ of the system.
We observe that in the Zubarev's seminal papers \cite{Zub}, an infinitesimal source term, corresponding to our $\epsilon$, was introduced \textit{ad hoc} in the construction of the nonequilibrium statistical operator from the LE, whereas, here, in the context of thermostatted particle systems, a small parameter arises naturally after the adimensionalization.
The parameter $\epsilon$ can be small or large, depending on the details of the coupling with the thermostat. It is well known that, in strongly dissipative systems ($\epsilon\gg 1$), the support of the invariant measure is an attractor endowed with a very thin fractal structure, which appears strongly at variance with a regular distribution. On the other hand, in the limit of weak dissipation (i.e.: weak coupling with the thermostat), $\epsilon\ll1$, it makes sense to attempt a perturbative technique to solve Eq. (\ref{Liouv}), qhich is performed by expanding $\rho$ in powers of $\epsilon$:

\be
\rho=\sum_{l=0}^{\infty}\epsilon^{l} \rho^{(l)}\label{expan}
\ee
where $\rho^{(0)}$ obeys a purely conservative equilibrium dynamics (unaffected by the external field and by the thermostat):

\be
\rho^{(0)}(t)=e^{-\Lambda t}\rho^{(0)}(0)
\ee
with $\Lambda=\dot{\Gamma_{0}}\cdot \nabla$ the Liouvillian operator. Let us point out, here, the analogy with Sec. \ref{sec:sec2}. In the context of the BE, we have already remarked that, in the limit $\epsilon \rightarrow \infty$, it holds $f \rightarrow f^{LM}$, i.e., the local equilibria (\ref{LM}) are a good approximation of the exact statistics of the system.
Similarly, we find, here, that for vanishing $\epsilon$, $\rho\rightarrow\rho^{(0)}$, i.e. the density reduces to the purely Hamiltonian contribution.\\
With (\ref{eom}) and (\ref{expan}), Eq. (\ref{Liouv}) transforms into:
\be
\dot{\rho}= -\dot{\Gamma_{0}} \cdot \nabla (\rho^{(0)}+\epsilon \rho^{(1)}+...)-\epsilon(R\cdot\nabla -\kappa)(\rho^{(0)}+\epsilon \rho^{(1)}+...)  \label{micro}
\ee

Moreover, if the assumption of time scales separation holds, one may wish to decompose the dynamics of the probability density into a contribution (expressed via a projection operator $P$, to be defined below in  (\ref{projector})) which is solely dependent on some relevant \textit{slow} variables, plus a contribution given in terms of an orthogonal projector $Q=1-P$ (with $1$ the identity operator), which features the dynamics of the \textit{fast} variables.
The decomposition of the dynamics of the density as induced by a specific choice of the projector $P$ was provided by Robertson \cite{Rob}:

\be
\dot{\rho}=P\dot{\rho}+\int_{0}^{t}K(t,t') \rho(t') dt' \label{Rober}
\ee

where the term $K(t,t')\rho(t')$ includes the contribution of fluctuations of the overwhelming majority of the \textit{fast} degrees of freedom and makes Eq. (\ref{Rober}) essentially nonlocal in time.
One may be tempted to investigate under which general conditions the memory kernel in (\ref{Rober}) becomes negligible.
The answer is that the full dynamics of the probability density $\dot{\rho}$, in (\ref{Liouv}), can be properly approximated by the projected part $P \dot{\rho}$ in (\ref{Rober}), for those systems characterized by a rapid decay of fluctuations of the fast degrees of freedom. This approximation, which is referred to in the literature as the \textit{adiabatic approximation} \cite{Just}, paves the way to determine a manifold of slow variables in the phase space. As already shown in previous works addressing the BE \cite{Matteo2}, the adiabatic approximation, which postulates a decomposition between fast and solw variables, has no immediate connection with the order of magnitude of the parameter $\epsilon$. In fact, the standard Chapman-Enskog technique, which relies on the adiabatic approximation in order to compare (\ref{microBE}) with (\ref{macroBE}), works properly just for $\epsilon \ll 1$; on the other hand, the method discussed in \cite{Matteo,Matteo2}, which also disregards the contribution of the dynamics of the fast variables in (\ref{Rober}), was shown to correspond to an exact summation of the Chapman-Enskog expansion and does not impose any constraint on the magnitude of $\epsilon$.
Now, in the spirit of the standard Chapman-Enskog technique, we may assume that the probability density $\rho$ is a sufficiently smooth function whose time dependence is  parameterized through a set of $N$ phase space functions $\langle A_{\mu}\rangle(t)$, with $\mu=1,...,N$.
We apply, then, the chain rule (summation over repeated indices is assumed) and write:

\be
P\dot{\rho}
=\sum_{j=0}^{\infty}\epsilon^{j}\frac{\delta \rho^{(j)}}{\delta \langle A_{\mu}\rangle} \partial_{t}\langle A_{\mu}\rangle \label{projector}
\ee

The dynamics of  $\langle A_{\mu}\rangle$ (provided that the boundary terms, in the integration by parts, are negligible) is found to be given by:
\be
\partial_{t}\langle A_{\mu}\rangle=\sum_{k=0}^{\infty}\epsilon^{k}\partial_{t}^{(k)}\langle A_{\mu}\rangle
\ee

where:
\be
\partial_{t}^{(k)}\langle A_{\mu}\rangle=\int \nabla A_{\mu}\cdot \dot{\Gamma} \rho^{(k)} d\Gamma= \langle \nabla A_{\mu}\cdot \dot{\Gamma}\rangle^{(k)}\label{defin}
\ee

with $\dot{\Gamma}$ given by (\ref{eom}).
Therefore, we obtain:
\be
P \dot{\rho}=\sum_{j,k=0}^{\infty}\epsilon^{(j+k)}\partial_{t}^{(k)}\rho^{(j)}  \label{macro}
\ee

with:
\be
\partial_{t}^{(k)}\rho^{(j)}=\frac{\delta \rho^{(j)}}{\delta \langle A_{\mu}\rangle}\partial_{t}^{(k)}\langle A_{\mu}\rangle
\ee

It is now time to specify the various observables $\langle A_{\mu}\rangle$. A crucial aspect of many nonequilibrium theories is, in fact, concerned with the definition of a proper set of coarse-grained variables triggering the dynamics of certain phenomena of interest.  In the context of thermostatted many-particle systems, we suggest to consider the average phase space contraction rate $\langle \kappa \rangle$.
The latter appear to be a promising candidate, as it is a dynamical quantity which, in the steady state, attains a constant value related, in the thermodynamic limit, to the steady entropy production of Irreversible Thermodynamics \cite{RonCoh}.
Then, the protocol we follow is to reconstruct the dynamics of the probability density by matching, at any order of $\epsilon$, the original \textit{microscopic} dynamics given by (\ref{micro}) with the \textit{projected} dynamics (\ref{macro}). Thus, we obtain the following equation:
\be
\sum_{j,k=0}^{\infty}\epsilon^{(j+k)}\partial_{t}^{(k)}\rho^{(j)} =-\sum_{l=0}^{\infty}\epsilon^{l} \dot{\Gamma_{0}} \cdot \nabla \rho^{(l)} -\sum_{m=0}^{\infty}\epsilon^{(m+1)}(R\cdot\nabla -\kappa)\rho^{(m)} \label {main}
\ee

Since $\rho^{(0)}$ evolves according to purely Hamiltonian dynamics, at the first order, $\epsilon^{1}$, Eq. (\ref{main}) gives:

\be
-\dot{\Gamma_{0}} \cdot \nabla\rho^{(1)}-(R\cdot\nabla - \kappa) \rho^{(0)}=\frac{\delta \rho^{(1)}}{\delta \langle \kappa\rangle}\langle \nabla \kappa \cdot \dot{\Gamma_{0}} \rangle^{(0)} \label{good}
\ee

Equation (\ref{good}) addresses and defines the first order deviation from the reference Hamiltonian dynamics, by taking into account the effect of the external field and the phase space contraction induced by the thermostat. A remark is in order when discussing the derivative of the probability density $\rho^{(k)}$ with respect to a generic variable $\langle A_{\mu} \rangle$. The evaluation of the average is, in principle, performed through the knowledge of the full probability density. This is avoided, in the Chapman-Enskog theory, by requiring that the collision invariants are defined only through the local Maxwellians (\ref{LM}). In the present context, instead, the assumption of weak coupling limit ($\epsilon\ll 1$) is used to justify the equivalence: $\langle A_{\mu} \rangle\simeq \sum_{i=0}^{k}\epsilon^{i}\langle A_{\mu} \rangle^{(i)}$.
It is worth to mention that a historically relevant approximate solution of (\ref{good}), is given by the \textit{quasi-equilibrium} (or maximum-entropy) approximation. The technique employs a maximization of the Gibbs entropy, $S(t)=\int \rho\log \rho d\Gamma$, under some given constraints, and works at full equilibrium, i.e. for $\epsilon=0$ and in the limit $t\rightarrow \infty$ (this is, in fact, a method to derive the canonical ensemble in equilibrium statistical mechanics, cf. Ref. \cite{Zwanzig}). Nevertheless, it is not obvious that the technique also applies for $\epsilon > 0$, although some authors \cite{Zub2, Rob} consider it a valid principle in general. Furthermore, a major drawback of this maximization method, outside equilibrium, is that the set of variables and the corresponding constraints are usually not known (typical candidates are the invariants of motion and quantities prescribed by the boundary conditions). We can avoid employing this approximation, as, due to the specific choice of variables we made, we can attempt to tackle analytically Eq. (\ref{good}).
In the limit $t\rightarrow \infty$ and for $\epsilon \ll1$, we seek a solution of (\ref{good}) taking the modified Gibbs form:

\be
\rho\simeq\rho^{(0)}+\epsilon \rho^{(1)}=e^{-[\beta H_{0}(\Gamma)-\epsilon \psi_{\kappa} \kappa(\Gamma)]} \label{rho}
\ee
where the $\psi_{\kappa}$'s are Lagrange multipliers depending on the value of the chosen parameter $\langle \kappa\rangle$.
Let us observe that the functional form of $\rho$ in (\ref{rho}) resembles the expression proposed, through a different method, also in \cite{MLJ}.
For the projected time derivative in (\ref{good}), one finds:

\be
\frac{\delta\rho^{(1)}}{\delta \langle \kappa\rangle}\langle \nabla \kappa \cdot \dot{\Gamma_{0}} \rangle^{(0)}\simeq \rho \kappa(\Gamma) \langle \nabla \kappa \cdot \dot{\Gamma_{0}} \rangle^{(0)} \frac{\partial \psi_{\kappa}}{\partial \langle \kappa\rangle^{(1)}} \label {1}
\ee

On the other hand, the computation of the microscopic time derivative, corresponding to the left hand side of (\ref{good}), gives:

\be
\dot{\Gamma_{0}} \cdot \nabla \rho^{(1)}= (\dot{\Gamma}_{0} \cdot\nabla \kappa )\rho \psi_{\kappa}\label{2}
\ee

Then, by inserting (\ref{1}) and (\ref{2}) into (\ref{good}), and by integrating over the whole phase space, we obtain a differential equation for the $\psi_{\kappa}$'s:

\be
\frac{\partial \psi_{\kappa}}{\partial \langle \kappa\rangle^{(1)}}\langle \kappa\rangle^{(1)}=-\psi_{\kappa}[1+ O(\epsilon)]
\ee

Therefore, up to the first order in $\epsilon$, the density in (\ref{rho}) attains the form:

\be
\rho\simeq e^{-[\beta H_{0}(\Gamma)-\epsilon \frac{\kappa(\Gamma)}{\langle \kappa \rangle^{(1)}}]} \label{soluz}
\ee

For small external fields, the expression (\ref{soluz}) can be further linearized in the nonequilibrium parameter $\epsilon$, which enables us to deduce the  nonequilibrium steady state ensemble average of a generic (smooth enough) observable $A$ for a weakly dissipative thermostatted particle system:

\be
\langle A\rangle \simeq \langle A \rangle^{eq}+ \epsilon \frac{\langle A \kappa \rangle^{eq}}{\langle \kappa \rangle^{(1)}} \label{NESS}
\ee

Eq. (\ref{NESS}) is our main result, showing that the steady state average of an observable $A$ comprises a first order nonequilibrium correction proportional to the equilibrium correlation function between phase space contraction rate and the observable itself. The response formula above can be interpreted as the adiabatic limit of the Green-Kubo formula \cite{CTXX}, such that the fast degrees of freedom have been projected out and the time integral of the equilibrium time-correlation function is replaced by an \textit{effective} equilibrium correlation function.
Eq. (\ref{NESS}) corresponds to a first order projection, on a manifold of slow variables, of the FDT for deterministic dissipative particle systems.

%\begin{figure}
%    \centering
%    \subfigure
%    {
%        \includegraphics[width=6cm, height=4cm]{Lorenz}
%       \label{Lorenz}
%    }
%    \subfigure
%    {
%        \includegraphics[width=6cm, height=4cm]{Lorenz2}
%        \label{Lorenz2}
%    }
%    \caption{\subref{Lorenz} Electrical conductivity $\sigma$ vs. electric field $E_x$.\textit{Black dots}: numerical results of Lloyd \textit{et al.} \cite{Lloyd}.\textit{Blue line}: adiabatic approximation of the Liouville Equation.\\  \subref{Lorenz2} Magnification of Fig. (\ref{Lorenz}).}
%\end{figure}

\begin{figure}
\begin{center}
        \includegraphics[width=6.8cm, height=4.7cm]{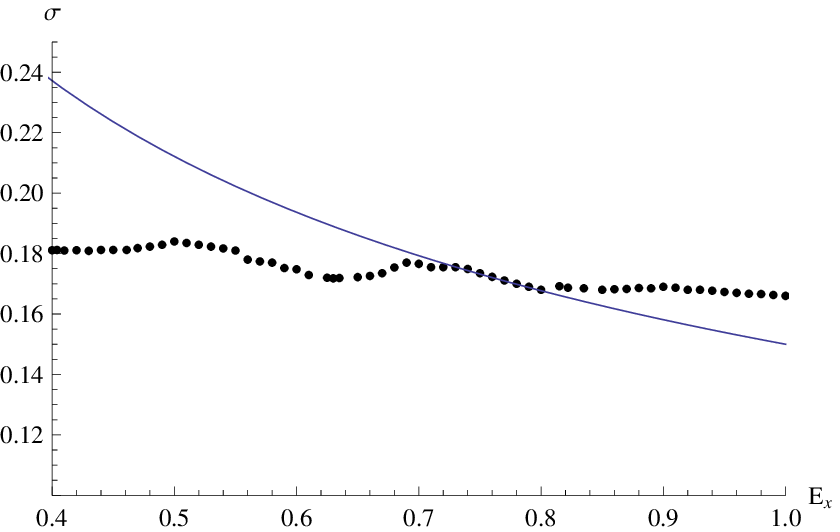}
        \caption{Electrical conductivity $\sigma$ vs. electric field $E_x$.\\ \textit{Black dots}: numerical results of Lloyd \textit{et al.} \cite{Lloyd}.\\ \textit{Blue line}: adiabatic approximation of the Liouville Equation}\label{Lorenz}
\end{center}
\hspace{0.7cm}
\begin{center}
        \includegraphics[width=6.8cm, height=4.7cm]{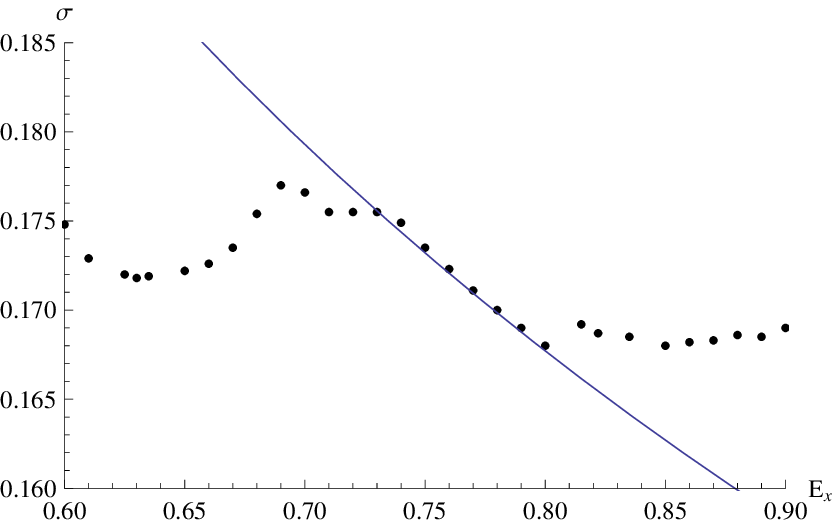}
        \caption{Magnification of Fig. (\ref{Lorenz}) in the range where the theory \\ best fits the numerical data.}\label{Lorenz2}
\end{center}
\end{figure}

By setting $A=\kappa$ and by recalling that $\textbf{E}=\sigma \textbf{j}$, with $\textbf{j}=q\langle \textbf{p}\rangle$, we used Eq. (\ref{NESS}) to compute, for instance, the electrical conductivity $\sigma(E_{x})$ of a Gaussian thermostatted periodic Lorentz gas driven by an external field $E_{x}$ parallel to the $x-$axis. The numerical analysis performed by Lloyd \textit{et al.} in Ref.\cite{Lloyd} shows that the $\sigma(E_{x})$ decreases nonlinearly with the field strength $E_{x}$, which could not be justified in terms of standard linear response theory and was understood as a typical nonlinear (higher order) effect. In turn, Eq. (\ref{NESS}) predicts, at the first order, the power law relation $\sigma \sim \frac{1}{\sqrt{E_{x}}}$ see Fig. (\ref{Lorenz}). Let us point out that our method applies, by construction, to dissipative systems, hence the external field must not vanish. Consequently, the agreement with experimental or numerical data is expected to be sensible in a range where the external field is finite and not exceedingly strong (as our result is pertinent to a first order perturbation theory). These expectations are, in fact, corroborated by the results shown in Fig. (\ref{Lorenz2}). This tends to suggest, \textit{a posteriori}, that, in nonequilibrium response theory for thermostatted particle systems, the adiabatic limit stands as a plausible physical approximation.

\section{Conclusions}

We remark that the novelty of the proposed approach stems from a suitable combination of a perturbation theory with a projection operator technique in the context of deterministically thermostatted particle systems with nonvanishing phase space contraction rate.
Our work aimed to show that the hypothesis of a time scales separation in a thermostatted particle system enables to express the expectation value of an observable solely in terms of few selected relevant variables (in our case the average phase space contraction rate) which drive the dynamics of the probability density.
The issue of determining response formulae in terms of few relevant observables is crucial to establish a bridge between a macroscopic hydrodynamic-like description and the underlying Liouville Equation.
Alternative derivations which lead to response formulae deduced from the exact Liouville propagator, without resorting to any projection technique, are known \cite{Evans}. We believe it is convenient to also look at approximate formulae which might offer a complementary point of view and might be, in some cases, of more immediate application, as we showed above in computing the conductivity of the Lorentz gas.
It is also worth to notice that, regardless of the intensity of the external field, for any finite field the steady state attractor has a lower dimension than the embedding space. Thus, since the support of the steady state measure lies on the attractor, the measure is not smooth and one has to resort to SRB measures \cite{Young}. On the other hand, the procedure of projecting out the fast degrees of freedom, is believed to ensure regularity to the resulting density, as also discussed in \cite{Evans2}.
Therefore, the suggested technique seems, also, to corroborate the applicability of the method proposed by Vulpiani \textit{et al.} in \cite{BPRVetc}, without invoking unstable manifolds of Anosov systems or introducing any noise occurring in not isolated physical systems
\ack
The Author acknowledges many inspiring discussions with Lamberto Rondoni and Wolfram Just and wishes, also, to thank Angelo Vulpiani and Rainer Klages for helpful comments.
This work was supported by the Swiss National Science Foundation (SNSF).

\end{document}